# A full vectorial model for pulse propagation in emerging waveguides with subwavelength structures part I: Kerr nonlinearity

**Shahraam Afshar V. and Tanya M. Monro**

*Centre of Expertise in Photonics, School of Chemistry & Physics, University of Adelaide, Adelaide, SA 5005, Australia*

*shahraam.afshar@adelaide.edu.au*

*http://www.physics.adelaide.edu.au/photonics*

**Abstract:** The propagation of pulses through waveguides with subwavelength features, inhomogeneous transverse structure, and high index contrast cannot be described accurately using existing models in the presence of nonlinear effects. Here we report the development of a generalised full vectorial model of nonlinear pulse propagation and demonstrate that, unlike the standard pulse propagation formulation, the z-component of guided modes plays a key role for these new structures, and results in generalised definitions of the nonlinear coefficient $\gamma$, $A_{eff}$, and mode orthognality. While new definitions reduce to standard definitions in some limits, significant differences are predicted, including a factor of $\sim 2$ higher value for $\gamma$, for emerging waveguides and microstructured fibers.

## References and links

## 1. Introduction

Nonlinear optical processes in optical fibers and waveguides have attracted significant interest because of the unique environment that they provide for nonlinear interactions, including tight confinement (high intensity), long interaction lengths, and control of propagation constants (see [1-3] and references therein). Recent and rapid progress in design and manufacturing of complex structured microstructured fibers and planar waveguides with subwavelength features (including both subwavelength inclusions and voids) has further extended the opportunities for guided-wave nonlinear optics and nonlinear devices by enabling extreme nonlinearity to combine with tailorable chromatic dispersion [1-4].

The nonlinear optical phenomena that occur in waveguides are determined through two main factors; the linear and nonlinear properties of the constituent bulk materials, and the optical properties of the waveguide. Two recent advances, as indicated below, have provided great potential to accelerate the field of guided-wave nonlinear optics: 1) the design and fabrication of complex structured waveguides with high contrast linear refractive indices and inhomogeneous cross sections, especially through postprocessing techniques. 2) the design and fabrication of waveguides with subwavelength features have opened up extensive opportunities for tailoring the nonlinear processes in waveguides.

While the characterisation of the nonlinear properties of bulk materials is indeed a rich and established field, the possibility of using postprocessing techniques to fill or coat complex structured waveguides with highly nonlinear materials has provided extended flexibilities in tailoring the nonlinear effects in waveguides and hence has opened new horizons for applications of nonlinear optical phenomena. The propagation of guided modes in waveguides with inhomogeneous cross sections are affected by their structure both directly, through the linear part of the refractive index which determines the modal characteristics, and indirectly, since the waveguide modes, under propagation through the structure, experience different losses, nonlinearity etc. Examples of such structures include liquid-filled [5-8] , gas-filled [9-13] , silicon nanocrystals-filled [14], atomic vapor-filled [15-18] , or surface-functionalised recently attracted significant interest, both in planar waveguides, [3,20-29] and fibers [30-33]. For such waveguides, it has been shown that a high intensity layer forms at the low refractive index side of the interface between two dielectric media due to the discontinuity of the normal component of the electric field. The intensity enhancement is proportional to the refractive index contrast of the two media. For the case of subwavelength voids in dielectrics, the enhanced intensity region forms at the surface of the void in the air side and extends over the whole void region since negligible evanescent decay of the field can occur within a subwavelength void. This key characteristic of the subwavelength features, when deployed within high intensity regions within the mode cross-section, can be used to achieve arbitrary distributions of high-intensity regions within waveguides, see for example [33]. We refer to this new class of optical waveguide, with high index contrast, inhomogeneous and complex structure, or subwavelength features, as 'Emerging waveguides' throughout this document.

Despite the importance and growing interest, applications, and publications in the field of nonlinear processes in these emerging waveguides, linear and nonlinear pulse propagation models for these structures still mainly rely on the well-known scalar Helmholtz equation [1,25,27,28,34-42]. This equation is based on the weak guidance approximation, and assumes that the waveguide cross-section is homogeneous structure. However, even a cursory inspection of the key characteristics of these emerging waveguides (i.e. inhomogeneous, high index contrast transverse structure incorporating subwavelength features) reveals that these waveguides operate far from the weak guidance regime. Indeed, one can argue that the emerging waveguides considered here exhibit strong guidance. For example, it is observed that no Helmholtz wave equation can be obtained for the emerging waveguides since $\nabla.D = 0$ in Maxwell's equation does not result in $\nabla.E = 0$ because of the inhomogeneous nature of susceptibility tensor $\varepsilon(x,y)$ [4]. Also, it has been pointed out that for subwavelength structures, such as optical nanowires, pulse propagation based on scalar theory does not give a good approximation [37].

There have been some reports of new models of nonlinear pulse propagation that take into account the inhomogeneous and vectorial solutions of Maxwell's equations [4,43-47]. However, Refs. [4, 43, 44, 45, 47] ignored the potential for coupling between different modes including different polarisations, which as we will show later, is a key characteristic of linear and nonlinear pulse propagation when the full vectorial solutions of Maxwell's equations are considered. In fact, we demonstrate here for the first time that there are parameter regimes for which this modal coupling makes a significant contribution to both the predicted nonlinear and dispersive effects. Also, in some reports, [43, 44, 46], no consideration is given to the possibility of using an inhomogeneous cross-section which, as mentioned above, is a key feature of emerging waveguides. In addition, one vital aspect of the vectorial formulation of nonlinear pulse propagation, which has not been considered to date to the best of our knowledge, is the impact of the longitudinal component of the modal fields [the component along the propagation direction, (z)] on the dispersion, nonlinear, and modal/polarisation coupling behaviour of a pulse propagating through an emerging waveguides. One reason why the z-component of the fields has not been considered before, especially in studies that report high nonlinearity in waveguides e.g., [4, 48], could be the fact that for slot waveguides TE modes, for which the z-component of the electric field is zero, have higher nonlinearity than those of TM modes.

Here a general vectorially-based Nonlinear Schrödinger Equation (VNSE), is derived for pulse propagation through waveguides with complex transverse structure including inhomogeneous refractive index profiles and subwavelength features. We demonstrate that in the strong guidance regime, the propagating modes have significant components along the direction of propagation, which causes the propagating modes to be non-transverse. As a result, this formalism predicts that a range of new tempo-spatial effects should be observable within emerging waveguides, such as dispersion-induced depolarisation.

Based on this VNSE, we derive a new and generalized equation for the $A_{eff}$, the parameter that defines the effective mode area, and $\gamma$, the parameter commonly used to describe the effective nonlinearity of an optical fiber [42]. These new definitions take into account both an inhomogeneous refractive index profile and subwavelength features. We apply these definitions to nanowires, and show that in some regimes, the value of $\gamma$ can be a factor of two higher than that obtained using the standard definition. We provide an analysis of the value of $\gamma$ predicted by this new generalised model. The new model also predicts new coupling terms between different modes or polarisations of propagating modes, which are due to non-transverse nature of the modes. Some preliminary results of the concept of the new model was presented in Ref. [49]. Here, we present the extension of our theoretical model and detailed results. We develop the theory of vectorially-based Nonlinear Schrödinger Equation (VNSE) in Sections 2, in which we derive new definitions for $\gamma$, $A_{eff}$, and mode orthogonality. We apply the new model to step index cylindrical waveguides and analyse and compare the results of the new model with those of the standard model in section 3. Concluding remarks are given in Section 4.

## 2. Theory

We start with Maxwell's equations for electric, magnetic and induced polarization fields, $\tilde{E}, \tilde{H}$, and $\tilde{P}$ in the Fourier domain as

$$\nabla \times \tilde{\mathbf{E}}(\mathbf{r}, \omega) = i\mu_0 \omega \tilde{\mathbf{H}}(\mathbf{r}, \omega) \tag{1}$$

$$\nabla \times \tilde{\mathbf{H}}(\mathbf{r}, \omega) = -i\varepsilon_0 \omega \tilde{\mathbf{E}}(\mathbf{r}, \omega) - i\omega \tilde{\mathbf{P}}(\mathbf{r}, \omega), \tag{2}$$

where the Fourier transformation is given by

$$F(\mathbf{r},t) = \frac{1}{2\pi} \int \tilde{F}(\mathbf{r},\omega) e^{-i\omega t} d\omega, \tag{3}$$

and $F = \mathbf{E}$, $\mathbf{H}$, or $\mathbf{P}$. By considering a perturbative expansion $\tilde{\mathbf{P}}(\mathbf{r}, \omega) = \sum_{n=1}^{\infty} \tilde{\mathbf{P}}^{(n)}(\mathbf{r}, \omega)$, where $(n)$ represent the order of induced polarization, $\tilde{\mathbf{P}}^{(1)}(\mathbf{r}, \omega) = \varepsilon_0 \chi^{(1)}(-\omega;\omega) \tilde{\mathbf{E}}(\mathbf{r}, \omega)$, and $\tilde{\mathbf{P}}_{NL}(\mathbf{r}, \omega) = \sum_{n=2}^{\infty} \tilde{\mathbf{P}}^{(n)}(\mathbf{r}, \omega)$ in which the second rank tensor $\chi^{(1)}(-\omega;\omega)$ is assumed to be a scalar and related to refractive index $n$ through $n^2(\mathbf{r},\omega) = 1 + \chi^{(1)}(-\omega;\omega)$, we find

$$\nabla \times \tilde{\mathbf{E}}(\mathbf{r}, \omega) = i\mu_0 \omega \tilde{\mathbf{H}}(\mathbf{r}, \omega) \tag{4}$$

$$\nabla \times \tilde{\mathbf{H}}(\mathbf{r}, \omega) = -i\varepsilon_0 n^2(\mathbf{r}, \omega) \tilde{\mathbf{E}}(\mathbf{r}, \omega) - i\omega \tilde{\mathbf{P}}_{NL}(\mathbf{r}, \omega). \tag{5}$$

Next we consider Eqs. (4) and (5) for two sets of fields; unperturbed fields $\tilde{E}_0(\mathbf{r}, \omega_0)$ and $\tilde{H}_0(\mathbf{r}, \omega_0)$, which represent the electromagnetic fields of narrowband pulses at frequency $\omega_0$ for which the dispersion, loss, and nonlinearity terms are zero, and perturbed fields $\tilde{E}(\mathbf{r}, \omega)$ and $\tilde{H}(\mathbf{r}, \omega)$, representing electromagnetic fields of frequency $\omega$ associated with wideband pulses centred at $\omega_0$, where the dispersion, loss and nonlinearity terms are nonzero. Vectorial solutions of Maxwell's equation for the unperturbed fields results in a complete orthonormal set of forward, backward and radiation propagating modes (labeled $\mu$) with propagating constants of $\beta_\mu$ and forward modal fields of; [50]

$$\widehat{\mathbf{e}}_\nu = \frac{\mathbf{e}_\nu(x,y,\omega_0)}{\sqrt{N_\nu}} e^{i\beta_\nu z} \tag{6}$$

$$\widehat{\mathbf{h}}_\nu = \frac{\mathbf{h}_\nu(x,y,\omega_0)}{\sqrt{N_\nu}} e^{i\beta_\nu z}, \tag{7}$$

where

$$\int \mathbf{e}_\mu(x,y,\omega) \times \mathbf{h}_\nu^*(x,y,\omega).\hat{z}dA = N_\mu \delta_{\mu\nu} \tag{8}$$

$$N_\mu = \frac{1}{2}\left|\int \mathbf{e}_\mu(x,y,\omega) \times \mathbf{h}_\mu^*(x,y,\omega).\hat{z}dA\right|.$$

The propagation constant $\beta_\nu$ and modal field distributions, $\mathbf{e}_\nu(x,y,\omega_0)$ and $\mathbf{h}_\nu(x,y,\omega_0)$, of propagating modes of a waveguide, in general, can be obtained through various numerical methods including Finite Element methods [51], the Multipole Method [52, 53], etc. By constructing a function $\mathbf{F}_C$ defined as

$$\mathbf{F}_C = \tilde{\mathbf{E}}_0 \times \tilde{\mathbf{H}}^* + \tilde{\mathbf{E}}^* \times \tilde{\mathbf{H}}_0$$

and using Eqs. (4) and (5) for perturbed and unperturbed fields, we find [50]

$$\nabla \cdot \mathbf{F}_C = -i\mu_0(\omega - \omega_0)\tilde{\mathbf{H}}^*.\tilde{\mathbf{H}}_0 - i\varepsilon_0[\omega n^2(\mathbf{r},\omega) - \omega_0 n^2(\mathbf{r},\omega_0)]\tilde{\mathbf{E}}^*.\tilde{\mathbf{E}}_0 + i\omega\tilde{\mathbf{E}}_0.\tilde{\mathbf{P}}_{NL}^*(\mathbf{r},\omega). \tag{9}$$

Next we expand the perturbed fields $\tilde{\mathbf{E}}$ and $\tilde{\mathbf{H}}$ according to the orthonormal and complete modal set of forward, backward, and radiation modes of the unperturbed field as [50]:

$$\tilde{\mathbf{E}}(\mathbf{r},\omega) = \sum_\mu \tilde{a}'_\mu(z,\omega)\frac{\mathbf{e}_\mu(x,y,\omega_0)}{\sqrt{N_\mu}}e^{i\beta_\mu z} + \tilde{a}'_{-\mu}(z,\omega)\frac{\mathbf{e}_{-\mu}(x,y,\omega_0)}{\sqrt{N_{-\mu}}}e^{-i\beta_\mu z} + \text{Radiation Modes}, \tag{10}$$

$$\tilde{\mathbf{H}}(\mathbf{r},\omega) = \sum_\mu \tilde{a}'_\mu(z,\omega)\frac{\mathbf{h}_\mu(x,y,\omega_0)}{\sqrt{N_\mu}}e^{i\beta_\mu z} + \tilde{a}'_{-\mu}(z,\omega)\frac{\mathbf{h}_{-\mu}(x,y,\omega_0)}{\sqrt{N_{-\mu}}}e^{-i\beta_\mu z} + \text{Radiation Modes}. \tag{11}$$

Here index $-\mu$ refers to backward propagating modes and both forward and backward modes are orthogonal to radiation modes. Here, we only consider unidirectional pulse propagation for which we neglect the back scattering of a forward propagating laser beam and the nonlinearity associated with it [43, 54]. This is not strictly true, especially for nonlinear and coupling processes where counter-propagating fields exist in the fiber. It can be shown that the back-scattered field affects the overall nonlinearity for the forward modes but we leave the full investigation of this effect to future publications. Therefore, we only consider the first term in Eqs. (10) and (11) for expanding perturbed fields and hence ignore the coupling between the unperturbed field with the backward and radiation modes of the perturbed field. This will be discussed further later in this section. Unlike other reports that consider the modal expansion only for the nonlinear term [4, 43, 45], we consider the modal expansion in Eqs. (10) and (11) for both dispersion and nonlinear effects. It should also be noted that the frequency dependence of perturbed fields is totally contained within the coefficients $\tilde{a}'_\mu(z,\omega)$. Assuming that the unperturbed fields $\tilde{\mathbf{E}}_0(\mathbf{r},\omega)$ and $\tilde{\mathbf{H}}_0(\mathbf{r},\omega)$ are one of the propagating modes (e.g., mode $\nu$) of unperturbed case i.e.,

$$\tilde{\mathbf{E}}_0(\mathbf{r},\omega_0) = \frac{\mathbf{e}_\nu(x,y,\omega_0)}{\sqrt{N_\nu}}e^{i\beta_\nu z} \tag{12}$$

$$\tilde{\mathbf{H}}_0(\mathbf{r},\omega_0) = \frac{\mathbf{h}_\nu(x,y,\omega_0)}{\sqrt{N_\nu}}e^{i\beta_\nu z}, \tag{13}$$

and using the reciprocal theorem [50]

$$\frac{\partial}{\partial z}\int \mathbf{F}_C.\hat{z}dA = \int \nabla.\mathbf{F}_C dA, \tag{14}$$

results in

$$\frac{\partial}{\partial z}\tilde{a}'_\nu(z,\omega) = \frac{1}{4}\sum_\mu [A_{\nu\mu}+B_{\nu\mu}]\tilde{a}'_\mu - \frac{i\omega e^{-i\beta_\nu z}}{4\sqrt{N_\nu}}\int \mathbf{e}^*_\nu.\tilde{\mathbf{P}}_{NL}(\mathbf{r},\omega)dA. \quad (15)$$

Here,

$$A_{\nu\mu} = \frac{i\mu_0 e^{-i(\beta_\nu-\beta_\mu)z}}{\sqrt{N_\nu N_\mu}}(\omega-\omega_0)\int \mathbf{h}_\mu.\mathbf{h}^*_\nu dA \quad (16)$$

$$B_{\nu\mu} = \frac{-i\varepsilon_0 e^{-i(\beta_\nu-\beta_\mu)z}}{\sqrt{N_\nu N_\mu}}\int [\omega n^2(x,y,\omega) - \omega_0 n^2(x,y,\omega_0)]\mathbf{e}_\mu.\mathbf{e}^*_\nu dA. \quad (17)$$

Eq. (15) is a general first order differential equation that describes the propagation of amplitudes of the coupling coefficient of the perturbed field based on unperturbed one. This equation is similar to those reported in Refs. [43, 46], except that the dispersion terms in Eq. (15) include the coupling between different modes.

Although, $\tilde{a}'_\nu s$ are the coefficients of the perturbed fields, Eq. (15) is exact in the sense that no perturbation has been considered for dispersion and nonlinearity and hence this equation can be applied to describe, in general, any nonlinear or dispersion-based processes in an optical waveguide. The first and the second terms on the right hand side of the equation represent the dispersion and nonlinearity, respectively. Next, we perform a Taylor series expansion for the dispersion term in Eq. (15), around $\omega_0$. Depending on the bandwidth of the pulse around $\omega_0$, higher orders in the Taylor series can be considered to achieve better approximation for dispersion terms. For some cases, such as supercontinuum generation where extra wideband pulses propagating along the waveguide, however, it may be more appropriate to work with Eq. (15) directly. We separate the sum over the modes in Eq. (15) into self and cross terms to find;

$$\frac{\partial}{\partial z}\tilde{a}'_\nu(z,\omega) = i\sum_{n=1}^\infty \frac{(\Delta\omega)^n}{n!}\beta_\nu^{(n)}\tilde{a}'_\nu + i\sum_{\mu\neq\nu}\sum_n \frac{(\Delta\omega)^n}{n!}\beta_{\nu\mu}^{(n)}\tilde{a}'_\mu \quad (18)$$

$$-\frac{i\omega e^{-i\beta_\nu z}}{4\sqrt{N_\nu}}\int \mathbf{e}^*_\nu.\tilde{\mathbf{P}}'_{NL}(\mathbf{r},\omega)dA$$

where

$$\beta_\nu^{(1)} = \frac{1}{4N_\nu}\int \left[\mu_0|\mathbf{h}_\nu|^2 + \varepsilon_0 \frac{\partial}{\partial\omega}(\omega n^2)|_{\omega=\omega_0}|\mathbf{e}_\nu|^2\right]dA \quad (19)$$

$$\beta_\nu^{(n)} = \frac{\partial^n}{\partial\omega^n}\beta_\nu^1 \quad (20)$$

$$\beta_{\nu\mu}^{(1)} = \frac{e^{-i(\beta_\nu-\beta_\mu)z}}{4\sqrt{N_\nu N_\mu}}\int \left[\mu_0\mathbf{h}_\mu.\mathbf{h}^*_\nu + \varepsilon_0\frac{\partial}{\partial\omega}(\omega n^2)|_{\omega=\omega_0}\mathbf{e}_\mu.\mathbf{e}^*_\nu\right]dA \quad (21)$$

$$\beta_{\nu\mu}^{(n)} = \frac{\partial^n}{\partial\omega^n}\beta_{\nu\mu}^1. \quad (22)$$

Here, to avoid confusion, superscripts (1) and (n) correspond to the first and higher order dispersion of the propagating modes, respectively, and subscripts $\mu$ label the mode number. It is straightforward to show that $\beta_\nu^{(1)}$ in Eq. (19) is in fact $\beta_\nu^{(1)} = 1/V_g$ where $V_g$ is the group velocity as given in Ref. [50]. One important aspect of Eq. (18), which has not been reported before, is the existence of cross dispersive terms $\beta_{\nu\mu}^{(1)}$, Eq. (21), and their derivatives, Eq. (22). Such terms can only become significant if they are phase matched, i.e., $\beta_\nu = \beta_\mu$, otherwise fast oscillations of the $e^{-i(\beta_\nu-\beta_\mu)z}$ average to a negligible value.

Equations (21) and (22) result in a new process when the $\mu$ and $\nu$ refer to the two polarisations of 1 and 2 of one mode. It is well known that waveguides with three or higher-fold symmetries are not birefringent [55] i.e., for these waveguides $\beta_1 = \beta_2$. In this case, the phase terms in Eqs. (21) and (22) are equal to unity and hence do not average to a negligible value as in the non phase matched case. The cross dispersive terms, $\beta_{12}^{(1)}$ and $\beta_{12}^{(n)}$ in Eqs. (21) and (22) are basically modifications to the group velocity $\beta_1^{(1)}$ and higher order dispersion terms $\beta_1^{(n)}$ of the polarisation 1. They have non-zero values which, as it will be shown later, is mainly due to the fact that in the strong guidance regime the dot product of the two polarisations of one mode i.e., $\mathbf{e}_1.\mathbf{e}_2^*$ and $\mathbf{h}_1.\mathbf{h}_2^*$ are non-zero because of strong z-component of the fields which results in non-transversality of the modes. This key finding is discussed in more detail later in this section. The physical consequence of this is dispersion-induced depolarisation of the guided mode, i.e., a polarised guided mode depolarises even if the incident beam is initially coupled perfectly to one of the polarisation axes of the waveguide. For instance, assuming that the incident beam is perfectly launched along the polarisation 1, then it can be deduced from Eq. (18) that the amplitude of the field along the polarisation 2, i.e., $\tilde{a}_2'$ inside the fibre grows through $\sum_n \frac{(\Delta \omega)^n}{n!} \beta_{21}^{(n)} \tilde{a}_1'$.

Next, we develop the time domain equivalent of Eq. (18). We multiply both sides of Eq. (18) by $e^{-i(\omega - \omega_0)t}$, integrate with respect to $\omega$, consider the following definitions

$$a_\nu'(z,t) \equiv 1/2 \left[ a_\nu(z,t) e^{-i\omega_0 t} + c.c \right]$$
$$P_{NL}'(\mathbf{r},t) = 1/2 \left[ P_{NL}(z,t) e^{-i\omega_0 t} + c.c \right],$$

where $a_\nu'(z,t)$ and $P_{NL}'(\mathbf{r},t)$ are the inverse Fourier transforms of $\tilde{a}_\nu'(z,\omega)$ and $\tilde{P}_{NL}'(\mathbf{r},\omega)$, respectively, to find

$$\frac{\partial}{\partial z} a_\nu(z,t) = i \sum_{n=1}^{\infty} \frac{(i\partial/\partial t)^n}{n!} \beta_\nu^{(n)} a_\nu + i \sum_{\mu \neq \nu} \sum_{n=1}^{\infty} \frac{(i\partial/\partial t)^n}{n!} \beta_{\nu\mu}^{(n)} a_\mu \qquad (23)$$
$$- i\omega_0 \frac{e^{-i\beta_\nu z}}{4\sqrt{N_\nu}} (1 + \tau_{shock} \partial/\partial t) \int \mathbf{e}_\nu^* . \mathbf{P}_{NL}(\mathbf{r},t) dA,$$

where $\tau_{shock} = i/\omega_0$. Equation (23) is a general equation that describes the nonlinear pulse propagation in the time domain. The first two terms on the right hand side of this equation describe the dispersion of a pulse propagating through a waveguide. The last term includes all the nonlinear effects and considers a shock term of $(1 + \tau_{shock} \partial/\partial t)$ which is responsible for self phase modulation and self steeping of the pulse. This term, in various forms, has been considered in many publications [34,38,40,55-61].

A few points should be noted here about the shock term: 1) similar to [43], it is naturally derived through the $(\partial/\partial t)\mathbf{P}_{NL}$ in the Maxwell's equations without any approximation of a second order time derivative in the scalar wave equation (Helmholtz equation) which is usually used to describe the nonlinear pulse propagation [34,55-57]. 2) The whole frequency dependence of the perturbed field is inherently included through the expansions Eq. (10) and (11) and is contained completely within $\tilde{a}_2'$ coefficients. Thus there is no need to include the frequency dependence of the propagating modes in the shock term, through $A_{eff}$, as has been done in Refs. [34, 38, 42, 56, 57]. 3) There is no dispersive term associated with $\chi^{(3)}$, i.e., $(\partial/\partial \omega)\chi^{(3)} = 0$ since, as we show in Appendix, assuming a delta function form for the response function of Kerr nonlinearity results in frequency independence of $\chi^{(3)}$. For the nonlinear term in Eq. (23), since $\chi^{(2)} = 0$ for isotropic medium such as glasses, we only consider the third order Kerr nonlinearity for which we approximate $\mathbf{P}_{NL}(\mathbf{r},t) \approx \mathbf{P}^{(3)}(\mathbf{r},t)$ and assume that the nonlinear response function can be expressed in terms of delta functions, see Appendix, and hence $\mathbf{P}^{(3)}(\mathbf{r},t)$, for Kerr nonlinearity, can be written as [63]

$$\mathbf{P}^{(3)}(\mathbf{r},t) = (3/4)\varepsilon_0 \chi^{(3)}(-\omega_0;\omega_0,\omega_0,-\omega_0)|\mathbf{E}(\mathbf{r},t)\mathbf{E}(\mathbf{r},t)\mathbf{E}^*(\mathbf{r},t), \qquad (24)$$

where $\chi^{(3)}$ is a rank four tensor and | indicates tensorial multiplication. Other third order nonlinear effects such as Raman scattering, for which the response function is not an instantaneous function of time will be a subject of future publications. Also it is assumed that $\mathbf{P}_{NL}(\mathbf{r},t) \approx \mathbf{P}^{(3)}(\mathbf{r},t)$ is a small perturbation compared to the linear induced polarisataion $\mathbf{P}_L = \varepsilon_0 \chi^{(1)}(-\omega;\omega)\tilde{\mathbf{E}}(\mathbf{r},\omega)$, and higher order nonlinear effects are negligible. This is usually a valid approximation at low intensity fields and typical optical glasses due to their relatively low nonlinear properties. However, for some materials such as semiconductor-doped glasses [63-67] and some organic materials such as paratoloune sulphonate (PTS) [65, 68] optical processes based on higher order nonlinearity can be observed, due to their higher order nonlinear susceptibilities, at moderate pulse intensity. For such materials, higher order terms must be considered in the nonlinear polarisation field $\tilde{\mathbf{P}}_{NL}(\mathbf{r},\omega) = \sum_{n=2}^{\infty} \tilde{\mathbf{P}}^{(n)}(\mathbf{r},\omega)$.

The components of $\chi^{(3)}$ depends on the class symmetry of the crystal. Silica glasses have isotropic crystal structure [42] and silicon crystal, which is usually used in waveguides, have m3m point-group symmetry [1]. For isotropic materials, it can be shown that among 81 elements of $\chi^{(3)}_{ijkl}$ ($i,j,k,l = x,y,z$) only 21 are nonzero, which depend on only three independent quantities [69] i.e.;

$$\chi^{(3)}_{ijkl} = \chi^{(3)}_{xxyy}\delta_{ij}\delta_{kl} + \chi^{(3)}_{xyxy}\delta_{ik}\delta_{jl} + \chi^{(3)}_{xyyx}\delta_{il}\delta_{jk}, \qquad (25)$$

where

$$\chi_{xxxx} = \chi_{yyyy} = \chi_{zzzz} = \chi^{(3)}_{xxyy} + \chi^{(3)}_{xyxy} + \chi^{(3)}_{xyyx}. \qquad (26)$$

Considering Eqs. (25), Eq. (24) can be written as;

$$P^{(3)}_i(\mathbf{r},t) = (3/4)\varepsilon_0 \left[ \sum_j \chi^{(3)}_{xxyy} |E_j|^2 E_i + \sum_j \chi^{(3)}_{xyxy} |E_j|^2 E_i + \sum_j \chi^{(3)}_{xyyx}(E_j)^2 E^*_i \right], \qquad (27)$$

where $i$ and $j$ refer to $x,y,z$. For Kerr nonlinearity with the choice of frequencies in Eq. (24), i.e., $\chi^{(3)}(-\omega_0;\omega_0,\omega_0,-\omega_0)$, the condition of permutation symmetry requires that $\chi^{(3)}_{xxyy} = \chi^{(3)}_{xyxy}$. The magnitude of the terms in the right hand side of Eq. (26) depends on the origin of the nonlinear term. In the case of silica and other glasses they are mainly nonresonant electronic origins for which $\chi^{(3)}_{xyxy} \approx \chi^{(3)}_{xyyx}$ [42, 69] and hence Eq. (27) can be simplified to

$$\mathbf{P}^{(3)}(\mathbf{r},t) = (1/2)\varepsilon_0 \chi^{(3)}_{xxxx}[(\mathbf{E}.\mathbf{E}^*)\mathbf{E} + (1/2)(\mathbf{E}.\mathbf{E})\mathbf{E}^*]. \qquad (28)$$

For silicon, however, the third order nonlinearity can be expressed based on four independent values as

$$\chi^{(3)}_{ijkl} = \chi^{(3)}_{xxyy}\delta_{ij}\delta_{kl} + \chi^{(3)}_{xyxy}\delta_{ik}\delta_{jl} + \chi^{(3)}_{xyyx}\delta_{il}\delta_{jk} + \chi_d \delta_{ijkl}, \qquad (29)$$

where $\chi_d \equiv \chi_{xxxx} - \chi_{xxyy} - \chi_{xyxy} - \chi_{xyyx}$ represent the nonlinearity isotropy. Similar to Silica, for the choice of frequencies in the third order susceptibility and photon energies $\hbar\omega$ well above $E_g$ $\chi_{xxyy}(-\omega;\omega,-\omega,\omega) = \chi_{xyyx}(-\omega;\omega,-\omega,\omega) \approx \chi_{xyxy}(-\omega;\omega,-\omega,\omega)$ [1]. As a result, $\chi_{ijkl}$ becomes

$$\chi^{(3)}_{ijkl} = \chi_{xxxx}[\frac{\rho}{3}(\delta_{ij}\delta_{kl} + \delta_{ik}\delta_{jl} + \delta_{il}\delta_{jk}) + (1-\rho)\delta_{ijkl}],$$

where $\rho \equiv 3\chi_{xxyy}/\chi_{xxxx}$ characterizes the nonlinear anisotropy and its value in the telecom band is real and close to 1.27 [1]. Using this, Eq. (24) can be written for silicon as

$$\mathbf{P}^{(3)}(\mathbf{r},t) = (\rho/2)\varepsilon_0 \chi^{(3)}_{xxxx}[(\mathbf{E}.\mathbf{E}^*)\mathbf{E} + (1/2)(\mathbf{E}.\mathbf{E})\mathbf{E}^*] + (3/4)\varepsilon_0(1-\rho)\chi^{(3)}_{xxxx}\mathbf{E}.\mathbf{E}.\mathbf{E}^*, \qquad (30)$$

where $\mathbf{E}.\mathbf{E}.\mathbf{E}^* \equiv \sum_i E_i E_i E_i^* \mathbf{v}_i$ ($\mathbf{v}_i$ is a Cartesian unit vector). Within the rest of this paper we ignore the last term of Eq. (30), which in fact affects the polarization dependence of nonlinear phenomena inside silicon waveguides, and thus Eq. (30) is the same as Eq. (28) except the factor $\rho$. Considering the expansion in Eq. (10) and using Eq. (28) we can evaluate the integrand in the last term of Eq. (23), i.e., $(1/\sqrt{N_\nu})e^{-i\beta_\nu z}\mathbf{e}_\nu^*.\mathbf{P}_{NL}(\mathbf{r},t)$ as;

$$(1/\sqrt{N_\nu})e^{-i\beta_\nu z}\mathbf{e}_\nu^*.\mathbf{P}_{NL}(\mathbf{r},t) = (1/2)\varepsilon_0 \chi^{(3)}_{xxxx} \sum_{\mu,\eta,\zeta} \tag{31}$$

$$[(1/\sqrt{N_\mu N_\eta N_\zeta N_\nu})a_\mu a_\eta^* a_\zeta (\mathbf{e}_\mu.\mathbf{e}_\eta^*)(\mathbf{e}_\nu^*.\mathbf{e}_\zeta)e^{-i(\beta_\nu-\beta_\mu+\beta_\eta-\beta_\zeta)z}$$
$$+ (1/2\sqrt{N_\mu N_\eta N_\zeta N_\nu})a_\mu a_\eta a_\zeta^* (\mathbf{e}_\mu.\mathbf{e}_\eta)(\mathbf{e}_\nu^*.\mathbf{e}_\zeta^*)e^{-i(\beta_\nu-\beta_\mu-\beta_\eta+\beta_\zeta)z}]$$

where, Greek indices $\mu,\nu,\eta,\zeta$ represent different modes of the waveguide. The terms on the right hand side of this equation, once integrated over the waveguide cross section, are overlap integrals representing how different propagating modes of the fiber couple to each other through the nonlinearity. Equation (31) can be expanded as sum of terms with and without phase terms as:

$$(1/\sqrt{N_\nu})e^{-i\beta_\nu z}\mathbf{e}_\nu^*.\mathbf{P}_{NL}(\mathbf{r},t) = (3/4)\varepsilon_0 \chi^{(3)}_{xxxx} \times \tag{32}$$

$$\{(\frac{|a_\nu|^2 a_\nu}{3N_\nu^2})\left[2|\mathbf{e}_\nu|^4 + |\mathbf{e}_\nu^2|^2\right]$$

$$+ \sum_{\mu\neq\nu} (\frac{2a_\nu |a_\mu|^2}{3\sqrt{N_\nu^2 N_\mu^2}}) \left[|\mathbf{e}_\nu.\mathbf{e}_\mu^*|^2 + |\mathbf{e}_\nu.\mathbf{e}_\mu|^2 + |\mathbf{e}_\nu|^2|\mathbf{e}_\mu|^2\right]$$

$$+ \sum_{\mu\neq\nu} (\frac{a_\mu^* a_\nu^2}{3\sqrt{N_\nu^3 N_\mu}}) \left[2|\mathbf{e}_\nu|^2(\mathbf{e}_\mu^*.\mathbf{e}_\nu) + (\mathbf{e}_\nu)^2(\mathbf{e}_\mu^*.\mathbf{e}_\nu^*)\right] e^{-i(\beta_\mu-\beta_\nu)z}$$

$$+ \sum_{\mu\neq\nu} (\frac{2a_\mu |a_\nu|^2}{3\sqrt{N_\nu^3 N_\mu}}) \left[2|\mathbf{e}_\nu|^2(\mathbf{e}_\mu.\mathbf{e}_\nu^*) + (\mathbf{e}_\nu^*)^2(\mathbf{e}_\mu.\mathbf{e}_\nu)\right] e^{-i(\beta_\nu-\beta_\mu)z}$$

$$+ \sum_{\mu\neq\nu} (\frac{|a_\mu|^2 a_\mu}{3\sqrt{N_\mu^3 N_\nu}}) \left[2|\mathbf{e}_\mu|^2(\mathbf{e}_\mu.\mathbf{e}_\nu^*) + (\mathbf{e}_\mu)^2(\mathbf{e}_\mu^*.\mathbf{e}_\nu^*)\right] e^{-i(\beta_\nu-\beta_\mu)z}$$

$$+ \sum_{\mu\neq\nu} (\frac{a_\mu^2 a_\nu^*}{3\sqrt{N_\nu^2 N_\mu^2}}) \left[2(\mathbf{e}_\mu.\mathbf{e}_\nu^*)^2 + (\mathbf{e}_\mu)^2(\mathbf{e}_\nu)^2\right] e^{-2i(\beta_\nu-\beta_\mu)z}$$

$$+ \sum_{\mu\neq\eta\neq\zeta\neq\nu} \text{other phase terms}\}.$$

The first two terms on the right hand side of Eq. (32) are automatically phase matched while the rest of the terms require phase matching in order to make significant contributions. The phase terms are responsible for nonlinear-induced depolarisation or four-wave-mixing [42]. They can be phased match, depending on $\Delta\beta_{\nu\mu} = \beta_\nu - \beta_\mu$. This can be achieved by employing the flexibility in controlling the dispersion properties of MOfs through structure design and glass choice.

By substituting Eq. (32) into Eq. (23), a first order differential equation is obtained which describes the nonlinear pulse propagation in a multimode waveguide. An important aspect of our formalism is related to the orthogonality of the waveguide propagating modes. Contrary to the standard formalism [42, 69, 70], for which $\mathbf{e}_\nu$s are approximated to be transverse modes and $\int \mathbf{e}_\nu^* \cdot \mathbf{e}_\mu dA = \int \mathbf{e}_{\nu t}^* \cdot \mathbf{e}_{\mu t} dA = 0$, or in the case of different polarizations $\mathbf{e}_\nu \cdot \mathbf{e}_\mu = 0$, in our formalism $\int \mathbf{e}_\mu^* \cdot \mathbf{e}_\nu dA \neq 0$ (or $\mathbf{e}_\nu \cdot \mathbf{e}_\mu \neq 0$ if $\mu$ and $\nu$ are the two polarizations of the same mode) because the modes are non-transverse, i.e., they have non-zero z-component. The generalized orthogonality condition, which is valid even in the strong guidance regime is $\int (\widehat{\mathbf{e}}_\nu \times \widehat{\mathbf{h}}_\mu^*) \cdot \hat{z} dA = \delta_{\nu\mu}$ inherently includes the $z-$component of the fields. Considering that

$$(\widehat{\mathbf{e}} \times \widehat{\mathbf{h}}^*) \cdot \hat{z} = (\widehat{\mathbf{e}}_t \times \widehat{\mathbf{h}}_t^*) \cdot \hat{z}, \tag{33}$$

and [50]

$$\widehat{\mathbf{h}}_t = (\frac{\varepsilon_0}{\mu_0})^{1/2} \frac{1}{k} \hat{z} \times [\beta \widehat{\mathbf{e}}_t + i \nabla_t \widehat{e}_z], \tag{34}$$

one can show that

$$\int (\widehat{\mathbf{e}}_\nu \times \widehat{\mathbf{h}}_\mu^*) \cdot \hat{z} dA = (\frac{\varepsilon_0}{\mu_0})^{1/2} \frac{1}{k} \int (\widehat{\mathbf{e}}_{\nu t} \times \{\hat{z} \times [\beta \widehat{\mathbf{e}}_{\mu t}^* - i \nabla_t e_{\mu z}^*]\}) \cdot \hat{z} dA, \tag{35}$$

$$= (\frac{\varepsilon_0}{\mu_0})^{1/2} \frac{1}{k} \int (\beta \widehat{\mathbf{e}}_{\nu t} \cdot \widehat{\mathbf{e}}_{\mu t}^* - i \widehat{\mathbf{e}}_{\nu t} \cdot \nabla_t \widehat{e}_{\mu z}^*) dA$$

which considering the general orthogonality relation $\int (\widehat{\mathbf{e}}_\nu \times \widehat{\mathbf{h}}_\mu^*) \cdot \hat{z} dA = \delta_{\nu\mu}$ results in

$$\int \widehat{\mathbf{e}}_{\nu t} \cdot \widehat{\mathbf{e}}_{\mu t}^* dA = (\frac{\mu_0}{\varepsilon_0})^{1/2} \frac{k}{\beta} \delta_{\nu\mu} + (\frac{i}{\beta}) \int (\widehat{\mathbf{e}}_{\nu t} \cdot \nabla_t \widehat{e}_{\mu z}^*) dA, \tag{36}$$

where subscript $t$ refers to transverse component of fields and operators. Eq. (36) clearly shows that $\int \widehat{\mathbf{e}}_{\nu t} \cdot \widehat{\mathbf{e}}_{\mu t}^* dA \neq 0$ in the parameter regime where $z-$component of electromagnetic fields are non zero. It should also be noted that Eq. (32) has been obtained by ignoring the backward and radiation terms in Eqs. (10) and (11). Considering these two terms in the expansion Eqs. (10) and (11) results in coupling between forward-backward and forward-radiation modes, which are represented by dot products of forward modes with backward and radiation modes in Eq. (32). These terms describe the power coupling between a forward propagating mode and backward and radiation modes due to nonlinearity. We leave the full investigation of these coupling effects to future publications.

In the case of single mode fibers, where two independent polarizations exist in the waveguide, $\mu$ and $\nu$ refer to the two polarizations 1 and 2 and $\Delta\beta_{\nu\mu}$ is the linear birefringence of the waveguide. For waveguides with strong birefringence, the beat length $L_B = 2\pi/\Delta\beta$ is short, and hence for fiber lengths $L >> L_B$ the phase terms oscillate very fast and hence have negligible contributions. However, for waveguides with weak birefringence, for which $L < L_B$ the phase terms are not negligible and should be taken into account. In the following sections we develop a model for nonlinear pulse propagation in single mode waveguides for both weak and strong birefringence.

### 2.1. Single mode highly birefringent waveguides

In the case of single mode waveguides with high birefringence, where only the first two terms in Eq. (32) are significant, substituting Eq. (32) into Eq. (23), considering [1, 42] $\varepsilon_0 c n^2 n_2 = (3/4) \operatorname{Re} \chi_{xxxx}^{(3)}$, where $n_2$ is the nonlinear refractive index of the material measured in $m/W$,

and ignoring nonlinear Two Photon Absorption we find:

$$\frac{\partial}{\partial z}a_\nu(z,t) = i\sum_n \frac{(i\partial/\partial t)^n}{n!}\beta_\nu^{(n)}a_\nu + \tag{37}$$
$$\frac{-ik}{4}(\frac{\varepsilon_0}{\mu_0})(1+\tau_{shock}\partial/\partial t)\{\frac{1}{3N_\nu^2}|a_\nu|^2 a_\nu \int n^2(x,y)n_2(x,y)\left[2|\mathbf{e}_\nu|^4+|\mathbf{e}_\nu^2|^2\right]dA$$
$$+ \frac{2}{3N_\nu N_\mu}|a_\mu|^2 a_\nu \int n^2(x,y)n_2(x,y)\left[|\mathbf{e}_\nu.\mathbf{e}_\mu^*|^2+|\mathbf{e}_\nu.\mathbf{e}_\mu|^2+|\mathbf{e}_\nu|^2|\mathbf{e}_\mu|^2\right]dA\},$$

where $\mu,\nu = 1,2$ and $\mu \neq \nu$ refer to the two polarisations of the fundamental mode. This equation can finally be written in a simple form;

$$\frac{\partial}{\partial z}a_\nu(z,t) = i\sum_n \frac{(i\partial/\partial t)^n}{n!}\beta_\nu^{(n)}a_\nu - \tag{38}$$
$$(1+\tau_{shock}\partial/\partial t)[i\gamma_\nu|a_\nu|^2 a_\nu + i\gamma_{\mu\nu}|a_\mu|^2 a_\nu],$$

where

$$\gamma_\nu = k(\frac{\varepsilon_0}{\mu_0})\frac{\int n^2(x,y)n_2(x,y)[2|\mathbf{e}_\nu|^4+|\mathbf{e}_\nu^2|^2]dA}{3|\int(\mathbf{e}_\nu \times \mathbf{h}_\nu^*).\hat{z}dA|^2}, \tag{39}$$

$$\gamma_{\mu\nu} = \gamma_{\mu\nu}^{(1)}+\gamma_{\mu\nu}^{(2)} = k(\frac{\varepsilon_0}{\mu_0})[(\frac{2\int n^2(x,y)n_2(x,y)\left[|\mathbf{e}_\nu.\mathbf{e}_\mu^*|^2+|\mathbf{e}_\nu.\mathbf{e}_\mu|^2\right]dA}{3|\int(\mathbf{e}_\mu \times \mathbf{h}_\mu^*).\hat{z}dA||\int(\mathbf{e}_\nu \times \mathbf{h}_\nu^*).\hat{z}dA|}+ \tag{40}$$
$$\frac{2\int n^2(x,y)n_2(x,y)|\mathbf{e}_\mu|^2|\mathbf{e}_\nu|^2 dA]}{3|\int(\mathbf{e}_\mu \times \mathbf{h}_\mu^*).\hat{z}dA||\int(\mathbf{e}_\nu \times \mathbf{h}_\nu^*).\hat{z}dA|}].$$

Eqs. (38) is the final form of nonlinear pulse propagation inside a single mode birefringent waveguide, which in form is similar to the commonly used equation (see [42]) but with the effective nonlinear coefficients of the waveguide are now given by the generalized forms as in Eq. (39) and (40). By generalizing the definition of the $A_{eff}$ as

$$A_{eff} = \frac{|\int(\mathbf{e}_\nu \times \mathbf{h}_\nu^*).\hat{z}dA|^2}{\int |(\mathbf{e}_\nu \times \mathbf{h}_\nu^*).\hat{z}|^2 dA} \tag{41}$$

the nonlinear coefficient $\gamma_\nu$ can be rewritten as

$$\gamma_\nu = \frac{2\pi}{\lambda}\frac{\overline{n_2}}{A_{eff}} \tag{42}$$

$$\overline{n_2} = (\frac{\varepsilon_0}{\mu_0})\frac{\int n^2(x,y)n_2(x,y)[2|\mathbf{e}_\nu|^4+|\mathbf{e}_\nu^2|^2]dA}{\int |(\mathbf{e}_\nu \times \mathbf{h}_\nu^*).\hat{z}|^2 dA},$$

where $\overline{n_2}$ can be viewed as nonlinear refractive index averaged over an inhomogeneous cross section weighted with respect to field distribution. The advantages of writing $\gamma$ as in Eq. (42) over the other reported form [4] is that it allows the analysis of $\gamma$ to be separated into parts describing linear (geometry and $n(x,y) \Longrightarrow A_{eff}$) and nonlinear (mode profile and $n_2(x,y) \Longrightarrow \overline{n_2}$) characteristics, providing a more intuitive analysis of $\gamma$. In our formalism, $A_{eff}$ has its standard interpretation as the effective area of the propagating modes, which can be determined purely based on the geometry and the linear refractive index of the waveguide $n(x,y)$, and does not require to be considered as the "effective nonlinear interaction area" as in Ref. [4]. Considering Eqs. (33) and (35), we find that Eq. (41) can be written as

$$A_{eff} = \frac{\left|\int [\beta |\mathbf{e}_t|^2 + i(\mathbf{e}_t.\nabla_t e_z)]dA\right|^2}{\int \left|\beta |\mathbf{e}_t|^2 + i(\mathbf{e}_t.\nabla_t e_z)\right|^2 dA}. \quad (43)$$

This equation in the limit of small $z$-component of electric field, where $\nabla_t e_z$ can be ignored in comparison with $\beta \mathbf{e}_t$, simplifies as

$$A_{eff} = \frac{\left|\int |\mathbf{e}_t|^2 dA\right|^2}{\int |\mathbf{e}_t|^4 dA}, \quad (44)$$

which is the standard definition of $A_{eff}$ [42].

Comparing Eqs. (38)-(42) with the expression commonly used for nonlinear birefringent terms [42], shows 1) the vectorial-based $\gamma$ developed here (referred to hereafter as $\gamma^V$) in Eq. (42) includes both the inhomogeneous waveguide structure and vectorial nature of electromagnetic fields and 2) an extra term that induces the depolarisation of an initially polarised beam through a non-zero coupling term $\gamma^{(1)}_{\mu\nu} \propto \int n^2(x,y) n_2(x,y) (|\mathbf{e}_v.\mathbf{e}^*_\mu|^2 + |\mathbf{e}_v.\mathbf{e}_\mu|^2) dA \neq 0$. The non-zero nature of this term, especially in the strong guidance regime as will be shown in Sec. 3, is the direct result of two facts; 1) the two different polarizations are not perpendicular to each other in the common sense, i.e., $\mathbf{e}_\mu.\mathbf{e}_\nu = 0$ or $\int \mathbf{e}_\mu.\mathbf{e}^*_\nu dA = 0$ because of the strong $z$-component of the fields and 2) the transverse integral, due to transverse dependence of $n^2(x,y)$ and $n_2(x,y)$, can be evaluated over different regions with different $n$ and $n_2$.

## 2.2. Single mode non-birefringent waveguides

For waveguides with perfect three or higher fold symmetry, the fundamental mode of the waveguide is degenerate or non-birefringent [55], i.e., $\beta_1 = \beta_2$ where 1 and 2 refer to the two polarisations. Therefore, the phase factors in Eq. (32) are equal to 1 for the pair of fundamental

modes of these waveguides and hence Eq. (32) and Eq. (23), result in :

$$\frac{\partial}{\partial z}a_\nu(z,t) = i\sum_{n=1}^{\infty}\frac{(i\partial/\partial t)^n}{n!}\beta_\nu^{(n)}a_\nu + i\sum_{n=1}^{\infty}\frac{(i\partial/\partial t)^n}{n!}\beta_{\nu\mu}^{(n)}a_\mu \quad (45)$$

$$-\frac{ik}{4}(\frac{\varepsilon_0}{\mu_0})(1+\tau_{shock}\partial/\partial t)\times \quad (46)$$

$$\{\frac{1}{3N_\nu^2}|a_\nu|^2 a_\nu \int n^2(x,y)n_2(x,y)\left[2|\mathbf{e}_\nu|^4 + |\mathbf{e}_\nu^2|^2\right]dA$$

$$+\frac{2}{3N_\nu N_\mu}|a_\mu|^2 a_\nu \int n^2(x,y)n_2(x,y)\left[|\mathbf{e}_\nu.\mathbf{e}_\mu^*|^2 + |\mathbf{e}_\nu.\mathbf{e}_\mu|^2 + |\mathbf{e}_\nu|^2|\mathbf{e}_\mu|^2\right]dA$$

$$+\frac{1}{3\sqrt{N_\nu^3 N_\mu}}a_\mu^* a_\nu^2 \int n^2(x,y)n_2(x,y)\left[2|\mathbf{e}_\nu|^2(\mathbf{e}_\mu^*.\mathbf{e}_\nu) + (\mathbf{e}_\nu)^2(\mathbf{e}_\mu^*.\mathbf{e}_\nu^*)\right]dA$$

$$+\frac{2}{3\sqrt{N_\nu^3 N_\mu}}a_\mu |a_\nu|^2 \int n^2(x,y)n_2(x,y)\left[2|\mathbf{e}_\nu|^2(\mathbf{e}_\mu.\mathbf{e}_\nu^*) + (\mathbf{e}_\nu^*)^2(\mathbf{e}_\mu.\mathbf{e}_\nu)\right]dA$$

$$+\frac{1}{3\sqrt{N_\mu^3 N_\nu}}|a_\mu|^2 a_\mu \int n^2(x,y)n_2(x,y)\left[2|\mathbf{e}_\mu|^2(\mathbf{e}_\mu.\mathbf{e}_\nu^*) + (\mathbf{e}_\mu)^2(\mathbf{e}_\mu^*.\mathbf{e}_\nu^*)\right]dA$$

$$+\frac{1}{3\sqrt{N_\nu^2 N_\mu^2}}a_\mu^2 a_\nu^* \int n^2(x,y)n_2(x,y)\left[2(\mathbf{e}_\mu.\mathbf{e}_\nu^*)^2 + (\mathbf{e}_\mu)^2(\mathbf{e}_\nu)^2\right]dA\},$$

where $\mu,\nu = 1,2$ and $\mu \neq \nu$ and refer to the two polarisations of the fundamental mode. Eq. (45) is the vectorial generalisation of the common nonlinear pulse propagation for the two polarisations of single mode fiber [42]. The main differences are the extra contributions from the different combinations of $\mathbf{e}_\mu.\mathbf{e}_\nu$. These terms, as indicated in the previous section, have nonzero values in the regime of high index and subwavelength core diameters.

### 3. Results and Discussion

The formalism developed in the previous section is general, and can be applied to an arbitrary waveguide. However, here we apply the above formalism to a simple step-index rod waveguide (i.e. a nanowire) and demonstrate that its nonlinear behaviour is predicted to be significantly different in the regime of high index contrast and subwavelength features than the predictions made using the standard formalism [42]. It should be pointed out that, throughout this paper we use the unit of $W^{-1}m^{-1}$ for $\gamma$ instead of the commonly used unit $W^{-1}km^{-1}$ [42]. This is justified considering the fact that small core structures in high index silicon or glasses can now provide access to waveguides with extremely large $\gamma$ values [4, 27, 39, 71].

A comparison between $\gamma^V$ developed here and the standard definition given by Agrawal [42];

$$\gamma^A = (2\pi n_2/\lambda)\frac{\int |F|^4 dA}{\left(\int |F|^2 dA\right)^2} \quad (47)$$

where $F(x,y) = e_t(x,y)$ is the scalar transverse electric field, indicates that $\gamma^V$ accounts for both inhomogeneous waveguide cross section and full vectorial nature of the propagating modes of the waveguide, especially the z-component of the modes. Figure 1 shows the effective nonlinearity for the two definitions of $\gamma$ as a function of core diameter for three step-index rods with different host materials of silica [42] ( $n = 1.45$, $n_2 = 2.6 \times 10^{-20}$ $m^2/W$), bismuth [72] ($n = 2.05$, $n_2 = 3.2 \times 10^{-19}$ $m^2/W$), and silicon [73] ($n = 3.45$, $n_2 = 4.5 \times 10^{-18}$ $m^2/W$ ).

While the $\gamma^V$ based on VNSE approaches $\gamma^A$ in the limit of large core diameter, it is significantly higher for small core diameters. For example, in Fig. 1c $\gamma^V$ is a factor of 2 higher than the $\gamma^A$ for silicon at the core diameter of $D = 0.19$ $\mu m$. Figure 1 also indicates that the difference between the $\gamma$ values increases as the index contrast of the core and cladding increases. Here, we have also considered another definition of $\gamma$, given by Foster et. al. [20], as;

$$\gamma^F = (2\pi/\lambda)\frac{\int n_2[(e_v \times h_v^*).\hat{z}]^2 dA}{[\int (e_v \times h_v^*).\hat{z} dA]^2}. \tag{48}$$

Foster et. al. [20] refer this equation to Agrawal [42]. It seems that Foster et. al. have just simply replaced $|F(x,y)|^2 = |e_t(x,y)|^2$ in Eq. (47) with the $S_z = (e_v \times h_v^*).\hat{z}$, arguing that $S_z$ is the intensity of light propagating down the waveguide. However, this replacement is only valid for the transverse mode approximation where $|F(x,y)|^2 = |e_t(x,y)|^2$ is proportional to $(e_v \times h_v^*).\hat{z}$ and interpreted as the intensity of the light, see Eq. (35). In general, for full vectorial formalism, the nonlinear-induced polarisation is expanded in terms of different powers of electric field strength, see theory section, which for the $\chi^{(3)}$ nonlinearity has the form given in Eq. (30), which in turn results in the new definition of $\gamma^V$ as in Eq. (39).

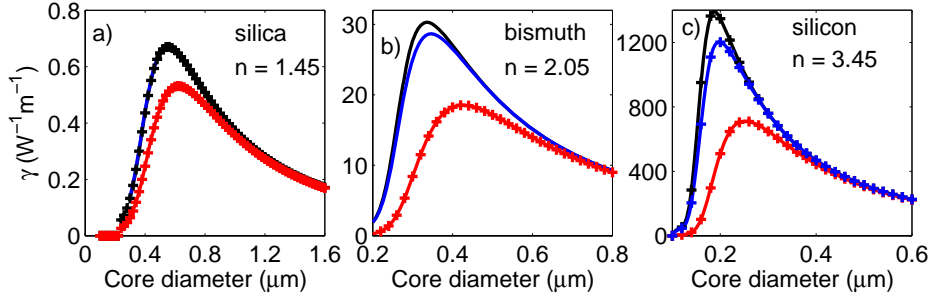

Fig. 1. Three definitions of $\gamma$, black is $\gamma^V$ based on VNSE, blue is $\gamma^F$ reported by Foster et. al., and red is $\gamma^A$ by Agrawal, as a function of core diameter and for three different materials silica ($n = 1.45$, $n_2 = 2.6 \times 10^{-20}$ $m^2/W$), bismuth ($n = 2.05$, $n_2 = 3.2 \times 10^{-19}$ $m^2/W$), and silicon ($n = 3.45$, $n_2 = 4.5 \times 10^{-18}$ $m^2/W$). The wavelength is $\lambda = 800$ $nm$, and the cladding is air with $n = 1.0$ in a) and b) and is silica, $n = 1.45$ in c). Plus signs and solid lines show the actual calculated data and the lines of best fit, respectively.

In Fig. 1, we have also plotted the values of $\gamma^F$ as a function of core diameter for different glass materials. While for silica, with the lowest index, $\gamma^V$ and $\gamma^F$ curves are on the top of each other for silicon with highest index, there is a maximum difference of 40% between the two curves at $D = 0.15$ $\mu m$. Also there is a difference of about 14% between the maximum values of $\gamma^V$ and $\gamma^F$.

The differences between $\gamma^V$ developed here and $\gamma^A$ and $\gamma^F$ are attributed to the fact that propagating modes of a waveguide are not transverse in strong guidance regime. In order to demonstrate this, we have defined the transversality [74] of a mode as $T_v = 1 - \int |e_{vz}^2| dA / \int |\mathbf{e}_v^2| dA$ and plotted it as a function of core diameter for different materials, as shown in Fig. 2. It indicates that in the regime of large core diameter, the transversality approaches 100%, i.e., the modes become essentially transverse as expected. However, in the limit of small cores, the transversality is reduced, indicating that a large fraction of electric field power is contained within the z-component of the field. This effect is more profound for the waveguides made from high index glass than the low index ones.

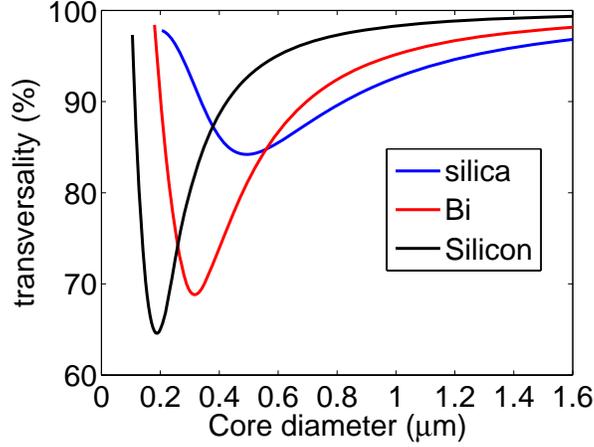

Fig. 2. Transversality versus core diameter for two glasses silica ($n = 1.45$) and bismuth ($n = 2.05$) and silicon ($n = 3.45$). The structure is a simple rod in the air for the glasses and a rod within the substrate of silica for silicon.

Fig. 3 shows a 2D plot of the z-component, $e_z$, and transverse electric field $\sqrt{e_r^2 + e_\theta^2}$, normalised to the power as in Eq. (6) for two different core diameters of $D = 0.4$ $\mu m$ and $D = 1.8$ $\mu m$ at the wavelength of $\lambda = 1550$ $nm$. It is evident from Figs. 3a and c that for large and small cores, the $e_z$ field is strongly localized at the edge of the fiber. For the core diameter $D = 0.4$ $\mu m$, however, the value of $e_z$ is one order of magnitude larger than that of the core diameter of $D = 1.8$ $\mu m$. Contrary to this, the distribution of the transverse field changes widely from a completely confined beam at large diameter, Fig. 3 d to a beam with high intensity regions at the fiber interface in Fig. 3b.

The first term in Eq. (40), $\gamma_{\mu\nu}^{(1)}$, is proportional to $\int n^2(x,y)n_2(x,y)(|\mathbf{e}_\nu.\mathbf{e}_\mu^*|^2 + |\mathbf{e}_\nu.\mathbf{e}_\mu|^2)dA$, which contributes to the overall nonlinearity of mode $\nu$, and is due to the overlap of the two different modes (two different polarizations in the case of single mode waveguide). This term does not appear in formalisms [42] where fully-transverse propagating modes and homogenous cross section are assumed since either $\int \mathbf{e}_\nu.\mathbf{e}_\mu^* dA = 0$, due to orthogonality of transverse modes or $\mathbf{e}_\nu.\mathbf{e}_\mu = 0$ if the modes are the two polarizations of a single fundamental mode.

In our formalism, however, this term appears because of the non-zero z-component of the fields which result in $\int |\mathbf{e}_\nu.\mathbf{e}_\mu^*|^2 dA \neq 0$ for any two propagating modes of a waveguide or $\mathbf{e}_\nu.\mathbf{e}_\mu \neq 0$ even when the two modes are the polarizations of the fundamental mode of a single mode waveguide. Figure 4 shows the magnitude of $\gamma_{\mu\nu}^{(1)}$ relative to $\gamma_\nu$, i.e., $\gamma_{\mu\nu}^{(1)}/\gamma_\nu$ where $\mu$ and $\nu$ are the two polarizations of a simple step index rod, as a function of core diameter. It demonstrates that indeed in the limit of large core, the relative value of $\gamma_{\mu\nu}^{(1)}$ approaches zero but for small core diameters its value is enhanced significantly. Comparing the ratio $\gamma_{\mu\nu}^{(1)}/\gamma_\nu$ for three different materials silica ($n = 1.45$) and bismuth ($n = 2.05$) and silicon ($n = 3.45$) demonstrates that the value of $\gamma_{\mu\nu}^{(1)}$ is more significant in subwavelength regime and for large index contrast host materials.

The behaviour of the nonlinear coefficient $\gamma^V$ as a function of wavelength also shows a sig-nificant difference between the usual definitions $\gamma^A$ and the one based on our VNSE. Fig. 5a) shows the behaviour of $\gamma$ as a function of core diameter for different wavelengths; $\lambda = 532$, 633, 800, 1064, 1310, and 1550 $nm$. As it is also evident from Fig. 5b, the $\gamma$ values decreases as the wavelength increases, but the decrease in the maximum value of $\gamma$ is much faster for the $\gamma^V$ of our model compared to that of common definition $\gamma^A$. It is also evident from Fig. 5a) that the position of maximum of $\gamma$ shifts to larger core diameters as the wavelength increases. Very high values of $\gamma$ at short wavelengths are due to the tighter confinement of the propagating mode of the waveguide and hence their higher intensities. The possibility of achieving $\gamma$ values of 150 $W^{-1}m^{-1}$, see Fig. 5b, suggest that order of $\pi$ nonlinear phase shift should be achievable for input powers of order of 20 $mW$ and for effective fibre length of 1 $m$ in the visible spectrum. The dispersion properties of the $\gamma$ values can be better compared by examining Fig. 6 in which the wavelength behaviour of $\gamma$ at constant core diameters are shown. While for large core diameter, e.g. $D = 1.6$ $\mu m$, the $\gamma$ ($\lambda = 1550$ $nm$) increases by a factor 3.5 to $\gamma$ ($\lambda = 532$ $nm$) and the two definitions of $\gamma$ are very close, for small core diameter $D = 0.5$ $\mu m$ the difference between the $\gamma$ values at $\lambda = 1550$ $nm$ and $\lambda = 532$ $nm$ is the order of 20 times and a large difference between the two definitions of $\gamma$ is observed. This indicates higher dispersion of $\gamma$ at small core diameters than that of large core diameters.

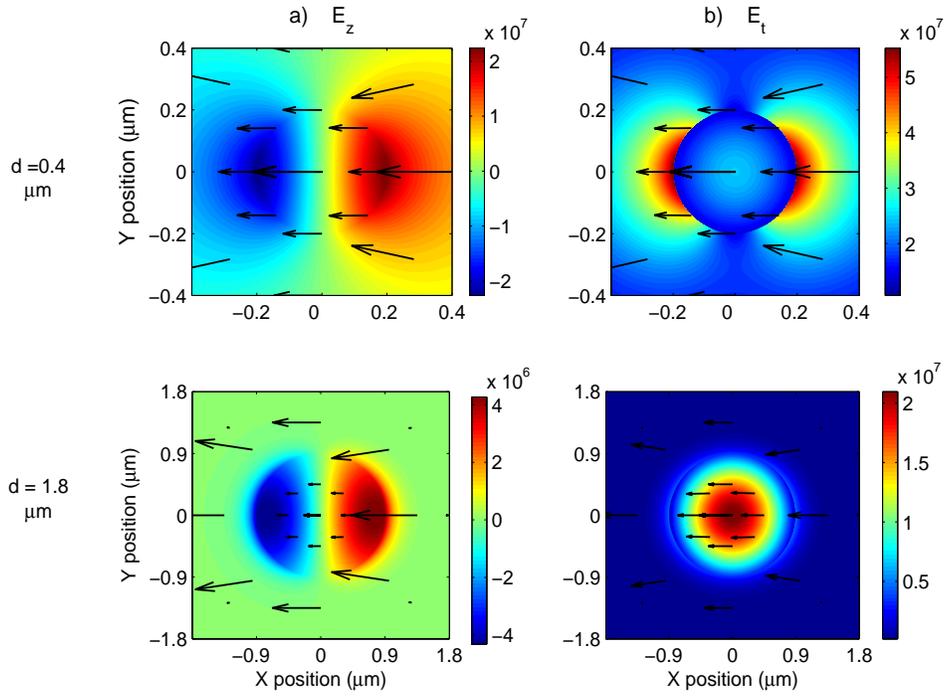

Fig. 3. 2D plot of $E_z$ (a,c) and $\sqrt{E_r^2 + E_\theta^2}$ (b,d) for two step index rods with core diameters 0.4 $\mu m$ (a,b) and 1.8 $\mu m$ (c,d) at the wavelength 1550 $nm$. The material is Bismuth with refractive index of $n = 2.05$.

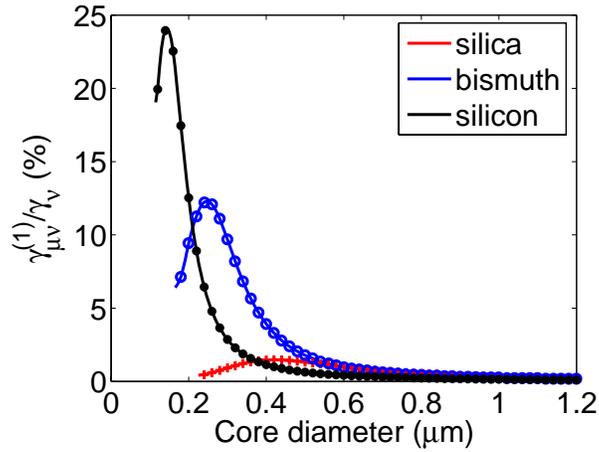

Fig. 4. Ratio of nonlinear coefficients $\gamma_{\mu\nu}^{(1)}$ and $\gamma_\nu$, see Eq. (40) and (39), as a function of core diameter for step index rods with host materials Silica ($n = 1.45$), Bismuth ($n = 2.05$), and Silicon ($n = 3.45$). The cladding material for glasses is air and for silicon is silica. Signs and the solid lines are the calculated data and lines of best fit, respectively.

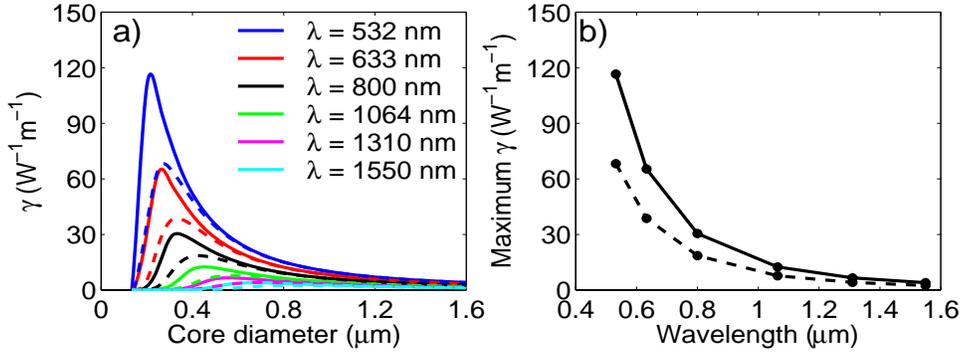

Fig. 5. a) $\gamma$ as a function of core diameter of a step index rod for different wavelengths. The solid lines are $\gamma^V$ and the dashed lines are $\gamma^A$. In b) the maximum of $\gamma$ in a) have been plotted as a function of wavelength. The host material is Bismuth ($n = 2.05$).

## 4. Discussion and conclusion

A new frontier in the field of optical waveguides is the design and fabrication of waveguides, referred to here as "emerging waveguides" with three main characteristics; 1) complex and inhomogeneous structure, 2) high index contrast, and 3) subwavelength features. A direct consequence of these features is that the common nonlinear Schrödinger equation which is based on weak guidance approximation does not provide accurate description of nonlinear processes in these waveguides. Here, we developed a vectorial based Nonlinear Schrödinger equation (VNSE), without relying on weak guidance approximation, which can be applied to these emerging waveguides.

An important feature of these waveguides is the fact that their propagating modes have much bigger z-component (z; direction of propagation) in comparison with those waveguides for which the weak guidance approximation is valid. As a result, the modes of these waveguides are not fully transverse and hence different orthogonality conditions govern them. Nonlinear and dispersion processes can be associated to this third "direction of propagation polarisation", which can also couples the transverse polarisations through dispersion and nonlinear processes. Our model provides a platform for generalising nonlinear processes such XPM, Modulation Instability, Soliton formation and propagation, Four Wave Mixing, Parametric Processes and Raman and Brillouin Scattering for emerging waveguides. The model also predicts new tempospatial processes such as dispersion-induced depolarisation of the guided modes. Despite the complexity that this new concept brings into guided-wave nonlinear optics, early redevelopment of Kerr and Raman processes indicates the great flexibility for "Engineering" nonlinear processes.

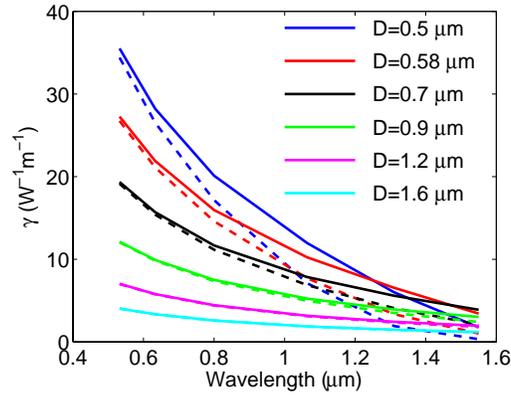

Fig. 6. $\gamma^V$ (solid lines) and $\gamma^A$ (dashed lines) vs wavelength for different core diameters for a step index rod with host material of Bismuth glass ($n = 2.05$).

Based on the model developed here, we provided a new vectorially-based definition for the effective nonlinear coefficient of waveguides ($\gamma^V$). Although the developed model is general, we have applied it to a simple step index cylindrical waveguide and shown that even for such a simple structure, $\gamma^V$ can be a factor of two higher than the common definition of $\gamma$, in the regime of high index contrast and subwavelength dimensions. However the full extent of the model in terms of exploring the rich physics behind the new pulse propagation model and the new definitions of effective nonlinearity $\gamma$, effective mode area $A_{eff}$, and cross-mode effective nonlinearity $\gamma_{\mu\nu}$, especially for waveguides with inhomogeneous structures, is yet to be explored.

The pulse propagation model developed here [see Eqs. (38) and (45)] adds some complexity in terms of numerical solutions in comparison with the standard model [42]. It includes calculations of different overlap integrals of the propagating fields and the linear and nonlinear index distribution, see Eqs. (38) and (45). These integrals, however, are numerically easy to take and need to be evaluated only once, for any given fibre, before numerically solving the pulse propagation equations. The model also implies that coupled pulse propagation equations of different modes, either different polarisations of the mode of a single mode waveguide or different modes of a multimode waveguide, must be solved to give an accurate picture of Kerr nonlinear process for a propagating pulse, especially in the parameter regime for which our formalism gives very distinct result in comparison with the standard model, i.e, high index contrast, inhomogeneous structure, and subwavelength features.

Experimental measurement of the effective nonlinearity in the parameter regime where there is a distinct difference between the new model and the standard model, will be crucial to confirm the results of new model. It should also pointed out that within this paper, Kerr nonlinearity has been considered for waveguides with inhomogeneous structure assuming instantaneous nonlinear response (i.e., response time much shorter than the pulse width $\tau_R \ll \tau_P$) of the materials everywhere in the waveguide. The case of finite and inhomogeneous nonlinear response time will be considered in future publications.

**Acknowledgements**

We acknowledge the Defence Science and Technology Organization (DSTO), Australia, for supporting research in the Centre of Expertise in Photonics. We would also like to acknowledge. This research was supported under Australian Research Council's Linkage Projects funding scheme (project number LP0776947). Tanya M. Monro acknowledges the support of an ARC Federation Fellowship.

**Appendix**

In general the third order induced polarisation is related to electric fields, in time and frequency domains, through the following equations [63];

$$\mathbf{P}^{(3)}(t) = \varepsilon_0 \int dt_1 dt_2 dt_3 R^{(3)}(t-t_1,t-t_2,t-t_3) |\mathbf{E}(t_1)\mathbf{E}(t_2)\mathbf{E}(t_3), \tag{49}$$

$$\mathbf{P}^{(3)}(\omega) = \varepsilon_0 \int d\omega_1 d\omega_2 d\omega_3 \chi^{(3)}(-\omega_\sigma;\omega_1,\omega_2,\omega_3) |\mathbf{E}(\omega_1)\mathbf{E}(\omega_2)\mathbf{E}(\omega_3) \delta(\omega-\omega_\sigma), \tag{50}$$

$$\chi^{(3)}(-\omega_\sigma;\omega_1,\omega_2,\omega_3) = \int dt_1 dt_2 dt_3 R^{(3)}(t_1,t_2,t_3) e^{i(\omega_1 t_1 + \omega_2 t_2 + \omega_3 t_3)}. \tag{51}$$

where $R^{(3)}$ is the third order response function. Assuming that for Kerr nonlinearity the response function is product of delta functions [42] i.e.,

$$R^{(3)}(t-t_1,t-t_2,t-t_3) = S^{(3)} \delta(t-t_1) \delta(t-t_2) \delta(t-t_3), \tag{52}$$

with $S^{(3)}$ being a constant, we find using Eq. (49);

$$P^{(3)}(t) = \int dt_1 dt_2 dt_3 R^{(3)}(t-t_1,t-t_2,t-t_3) |E(t_1)E(t_2)E(t_3), \tag{53}$$
$$= S^{(3)} |E(t)E(t)E(t).$$

By taking the Fourier transform of Eq. (53), one can deduce that third order polarisation in the frequency domain as

$$P^{(3)}(\omega) = \frac{S^{(3)}}{2\pi} \int d\tau E(\tau)E(\tau)E(\tau) e^{i\omega\tau} \tag{54}$$
$$= \frac{S^{(3)}}{2\pi} \int d\tau e^{i\omega\tau} \int d\omega_1 E(\omega_1) e^{-i\omega_1 \tau} \int d\omega_2 E(\omega_2) e^{-i\omega_2 \tau} \int d\omega_3 E(\omega_3) e^{-i\omega_3 \tau}$$
$$= \frac{S^{(3)}}{2\pi} \int d\omega_1 d\omega_2 d\omega_3 E(\omega_1)E(\omega_2)E(\omega_3) \delta(\omega-\omega_\sigma),$$

where $\omega_\sigma = \omega_1 + \omega_2 + \omega_3$. Equation (54) must be equivalent to Eq. (50) and hence we find that

$$\frac{S^{(3)}}{2\pi} = \varepsilon_0 \chi^{(3)}(-\omega_\sigma;\omega_1,\omega_2,\omega_3),$$

which means that $\chi^{(3)}(-\omega;\omega,-\omega,\omega)$ of the Kerr nonlinearity is not a function of frequency.